# Mechanical unfolding of a homopolymer globule: applied force vs applied deformation

*Alexey A. Polotsky,\* [1] Elizaveta E. Smolyakova [2], Oleg V. Borisov,[1, 3] and Tatiana M. Birshtein [1]*

[1] Institute of Macromolecular Compounds, Russian Academy of Sciences, Saint-Petersburg, Russia; E-mail: alexey.polotsky@gmail.com
[2] Saint-Petersburg State University, Physical Faculty, Saint-Petersburg, Russia
[3] Institut Pluridisciplinaire de Recherche sur l'Environnementet les Matériaux (IPREM) CNRS/UPPA, Pau, France

Summary : We propose the quantitative mean-field theory of mechanical unfolding of a globule formed by long flexible homopolymer chain collapsed in poor solvent and subjected to an extensional force We show that with an increase in the applied force the globule unfolds as a whole without formation of an intermediate state. The value of the threshold force and the corresponding jump in the distance between chain ends increase with a deterioration of the solvent quality and / or with an increase in the degree of polymerization. This way of globule unfolding is compared with that in the *D*-ensemble, when the distance between chain ends is imposed.

## Introduction

Experiments on single molecules subjected to extensional force (or to extensional deformation) have become possible recently due to development of such techniques as AFM single molecule force spectroscopy and optical or magnetic tweezers. These techniques allow operating on the length scale of nanometers and to measure (or to apply) forces in pN range and are now widely used, in particular in studies of biological macromolecules.

There exist two modes of micromechanical action on a macromolecule, either by applying a force *f* to its ends and measuring the average distance between them or by imposing the end-to-end distance *D* and measuring the average force of reaction on this deformation [1]. From the point of view of the statistical mechanics, one can consider these two situations as corresponding to different *ensembles*: constant force, or *f*-ensemble, and constant extension, or *D*-ensemble, respectively.

The present work is devoted to theoretical study of a system not as complicated as proteins or nucleic acids. Namely, we consider unfolding of a single homopolymer globule. Starting from the pioneering work of Halperin and Zhulina [2] in 1991, this problem has attracted a lot of attention of theorists (see [3 – 6] and references therein). In our resent work [7] we have performed an extensive self-consistent field modeling of globule unfolding in the *D*-ensemble in a wide range of polymerization degree and solvent quality. We have carried out a detailed analysis of conformations adopted by the chain with imposed end-to-end distance and calculated force-extension, *f(D)*, curves. Later on, in [8] we have proposed a simple analytical model and developed a quantitative theory which shown a very good agreement with the results obtained by SCF modeling and allowed to study in detail the intramolecular transitions.

The aim of the present work is to extend the approach developed in [8] to the case of the globule deformation in the *f*-ensemble and to compare the globule unfolding in *f*- and *D*-ensembles.

## Model and formalism

Consider a single flexible polymer chain consisting of $N$ (spherically symmetric) monomer units of size $a$ immersed into a poor solvent. Under these conditions the chain collapses and forms a spherical globule, Fig. 1 a. If we assume that the polymer density $\varphi$ within the globule is constant, then the radius of the (free) globule is $R_0 = [3N/(4\pi\varphi)]^{1/3}$.

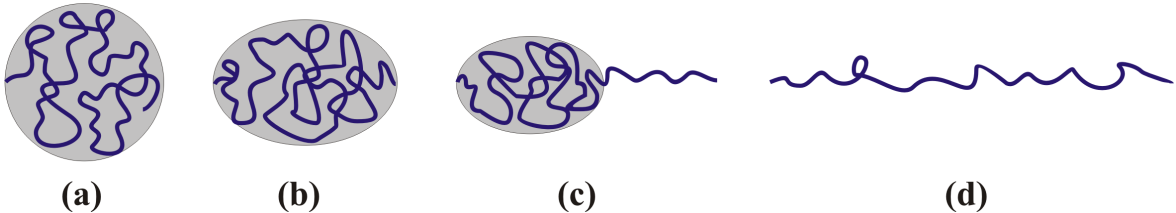

**(a)**      **(b)**      **(c)**      **(d)**

Fig. 1. Schematic pictures of free globule (a), prolate globule (b), tadpole (c), and stretched chain (d) conformations.

As it was shown in the previous studies [2 - 4, 7], when the globule is subjected to an external extensional deformation (i.e. in the *D*-emsemble), it can exist in one of conformational states schematically shown in Fig. 1. A weakly perturbed globule is still compact but has an elongated shape, Fig. 1 b, whereas if the force or deformation is strong the globule unfolds and can be represented as a stretched "open" chain, Fig. 1 d. An intermediate situation is also possible, in this case both compact and strongly stretched phases coexist in equilibrium in the molecule and it acquires the "tadpole" shape, Fig. 1 c.

Our approach consists in calculating and then comparing the free energies of the possible conformational states of the deformed globule shown in Fig. 1. As it was noted in the Introduction, there exists two ways of globule unfolding, either by imposing the end-to-end distance or by applying a pulling force. Depending on the ensemble, a proper thermodynamic potential (free energy) should be chosen. For the *D*-ensemble one should calculate the Helmholz free energy *F*, whereas in the *f*-ensemble one should work with the Gibbs free energy *G*. These thermodynamic potentials are related via Legendre transform:

$$G = F - Df, \qquad (1)$$

with $f > 0$ corresponds to an external pulling force in the *f*-ensemble or to the reaction force in the *D*-ensemble (oppositely directed to the pulling). Let us now calculate the Helmholz and the Gibbs free energies for the possible states of the deformed globule.

**Weakly deformed globule**

In this situation, Fig. 1 b, the globule can be modeled by a prolate ellipsoid with constant density. Let us first consider the case of the *D*-ensemble when the globule is deformed by separating the chain ends at the distance *D* and calculate the Helmholz free energy $F_{globule}$.

The main contributions to the free energy of the deformed globule are given by the volume and the interfacial contributions. The former is proportional to the degree of polymerization, the latter - to the ellipsoid surface area : $F_{globule} = \mu N + \gamma S$, where μ is the monomer chemical potential (free energy) in the infinite globule (μ is negative, thus reflecting poor solvent conditions) and γ is the interfacial tension coefficient (γ is positive due to the energetic and conformational penalty of redundant monomer-solvent contacts at the interface). Here and below $k_B T$ is taken as energetic unit. The volume of a prolate ellipsoid is $V = \pi D b^2 / 6$ where *D* and *b* are the major and minor axes, respectively. The ellipsoid surface area is given by $S = \frac{1}{2} \pi D b [\sqrt{1-\alpha^2} + (1/\alpha) \cdot \arcsin \alpha]$, where $\alpha = \sqrt{1-(b/D)^2}$ is the ellipticity parameter. Since the volume of the globule is conserved upon its deformation from spherical to ellipsoidal shape, i.e. $V = 4\pi R_0^3/3$, this leads to the following expression for the free energy

$$F_{globule} = \mu N + 4 \pi R_0^2 \gamma g(x), \qquad (2)$$

where $g(x)$ is a universal function of the parameter $x := D/(2R_0)$:

$$g(x) = \frac{1}{2x} + \frac{x^2}{2\sqrt{x^3-1}} \arcsin\sqrt{1-\frac{1}{x^3}} \; . \tag{3}$$

The corresponding reaction force $f$ is given by:

$$f_{globule} = \frac{\partial F_{globule}}{\partial D} = 2\pi R_0 \gamma \, g'(x) \; , \tag{4}$$

where the derivative of $g(x)$ is

$$g'(x) = \frac{1}{2}\left[ -\frac{1}{x^2} + \frac{3x}{2(x^3-1)} + \frac{x(x^3-4)}{2(x^3-1)^{3/2}} \arcsin\sqrt{1-\frac{1}{x^3}} \right] \; . \tag{5}$$

At small deformations, $D/Na << 1$, $F_{globule} \approx \mu N + \gamma\,[4\pi R_0^2 + 2\pi\,(D-2R_0)^2/5]$, from what follows that the reaction force $f$ grows linearly with extension: $f_{globule} \approx 4\pi\,\gamma\,(D-2R_0)/5$ .

In order to switch to the $f$-ensemble, we use the Legendre transform (1) and write down the Gibbs free energy:

$$G_{globule} = \mu\,N + 4\,\pi\,R_0^2\,\gamma\,g(x) - f\,D \; . \tag{6}$$

Expression (6) still contains the end-to-end distance $D$ as a parameter. Therefore $G_{globule}$ should be minimized with respect to $D$: $\partial G_{globule}/\partial D = 0$. This leads to the equation

$$2\,\pi\,R_0\,\gamma\,g'(x) = f, \tag{7}$$

where $g'(x)$ is given by Eq. (5). Analysis of Eq. (5) shows that $g'(x)$ is a nonmonotonic function. At small $x$ $g'(x)$ grows, at $x = x^* \approx 2{:}1942$ it passes through a maximum, $g^* := g'(x^*) \approx 0.2214$, and then decreases with extension. This means that Eq. (7) has a solution only if

$$f \leq f^* := 2\,\pi\,R_0\,\gamma\,g^*, \tag{8}$$

i.e. at $f > f^*$ the compact globular conformation is unstable upon action of a constant force. $f^*$ is therefore a spinodal point for the compact globular state.

Inserting the equilibrium value of $D$ we get the equilibrium free energy $G_{globule}$ as a function of applied force $f$. The approximate expression for the Gibbs free energy at $fa << 1$ is $G_{globule} \approx \mu N + 4\pi\gamma R_0^2 - 2R_0 f - 5f^2/(8\pi\gamma)]$ .

**Stretched ("open") chain**

At strong deformations the globule is completely unfolded and strongly stretched, Fig. 1 d. The conformational free energy of the stretched chain can be easily calculated if we consider an ideal chain extended by an external force (external field), i.e. in the $f$- ensemble.

Suppose that the chain walking on the cylindrical lattice is subjected to a force $f$ directed along the $z$-axis. Let the probability to make a step either in $r$ or in $z$ direction is given by $\lambda_1$, the probability of a step in "$rz$"-direction, i.e. simultaneously changing both r and z coordinates by $\lambda_2$, whereas the rest, $\lambda_0 = 1 - 4\lambda_1 - 4\lambda_2$ is the probability to change the angular coordinate $\varphi$. Then the partition function of a monomer is given by

$$w = (\lambda_0 + 2\lambda_1)e^0 + (\lambda_1 + 2\lambda_2)e^{fa} + (\lambda_1 + 2\lambda_2)e^{-fa} = 1 + \frac{\cosh(fa)-1}{2k}, \quad (9)$$

where $k = 1/(4\lambda_1 + 8\lambda_2)$. Then the partition function of the chain is $Z = w^N$ and the logarithm of the partition function gives us the Gibbs free energy

$$G_{chain} = -\log Z = -N \log\left[1 + \frac{\cosh(fa)-1}{2k}\right]. \quad (10)$$

Once the partition function or the Gibbs free energy is known, the extension $D$ [or the degree of extension $D/(Na)$] corresponding to the force $f$ can be immediately found:

$$D = -\frac{\partial G_{chain}}{\partial f} = Na \cdot \frac{\sinh(fa)-1}{2k + \cosh(fa)-1}. \quad (11)$$

In order to obtain the Helmholz free energiey for the fixed extension ensemble, we use the Legendre transform (1): $F_{chain} = G_{chain} + D\,f$ and the relation (11) between $D$ and $f$. A simple form of thermodynamic functions' dependences on the extension $D$ can be obtained in the $fa \ll 1$ limit. Then from Eqs. (10)-(11): $F_{chain} \approx kD^2/(Na^2)$, $G_{chain} \approx -4Nf^2a^2/k$, and $f_{chain} \approx 2kD/(Na^2)$, i.e. the elastic free energy has a Gaussian form and the force-extension dependence is linear in this regime. The physical meaning of the parameter $k$ becomes also clear from the simplified expressions: $k$ plays the role of elastic constant.

**Tadpole conformation**

At moderate extensions, a microphase segregation within the globule occurs, and the globule acquires a tadpole conformation with co-existing globular head and stretched tail, Fig. 1 c.

Consider first the $D$-ensemble. In order to describe the tadpole state, we assume that the ellipsoidal head comprises $n$ monomers and has the major semi-axis length equal to $d$. Therefore, the tail consists of $N-n$ monomers and stretched at the distance $D-d$. Then the Helmholz free energy of the tadpole can be written as

$$F_{tadpole} = F_{globule}(n, d) + F_{chain}(N - n, D - d) \quad (12)$$

The tadpole free energy (12) should be minimized with respect to $n$ and $d$ at each given value of $D$. It is straightforward to show that the minimization conditions are equivalent to equalities of chemical potentials of monomer unit reaction forces in globular and stretched phases, respectively :

$$\begin{cases} \mu_{globule}(n,d) = \mu_{chain}(N-n, D-d) \\ f_{globule}(n,d) = f_{chain}(N-n, D-d) \end{cases} \quad (13)$$

Phase equilibrium conditions, Eq. (13), allow obtaining all equilibrium characteristics of the two-phase system. Inserting the equilibrium values of $n$ and $d$ into $F_{tadpole}$, Eq. (12), gives the equilibrium free energy. Comparison of the latter with the free energies of one-phase states (i.e. with $F_{globule}$ and $F_{chain}$) gives the lowest free energy corresponding to globally equilibrium state as well as boundary values of extension $D$ where the transition between conformations occur.

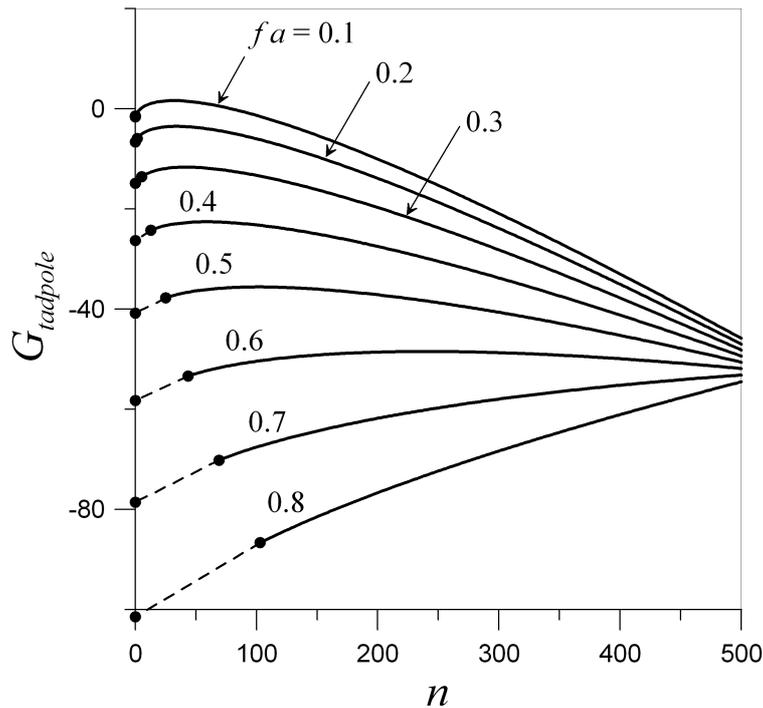

Fig. 2. Gibbs free energy of the tadpole conformation as function of the number of monomers in the globular head calculated for $N = 500$, $\chi = 1$ at various values of applied force $f$. Note also that the curves have a gap between $n \geq N$ and $n = 0$ because according to the condition (8) the globular head containing $n$ monomers is stable upon action of the force $f$ only if $n \geq N^* = f^3 \varphi /(6 \pi^2 \gamma^3 g^{*3})$.

For the $f$-ensemble we can use a similar reasoning noting that in this case the situation is even simpler than in the *D*-ensemble because the force balance condition [the second equation in (13)] is fulfilled automatically. The Gibbs free energy of the tadpole conformation in the *f*-ensemble reads

$$G_{tadpole} = G_{globule}(n) + G_{chain}(N - n) \qquad (14)$$

Then $G_{tadpole}$ should be minimized with respect to $n$. However, if we analyze $G_{tadpole}(n)$ dependence for fixed force, we see that the corresponding curve is convex up and $G_{tadpole}(n)$ has only boundary minima (i.e. at $n = 0$ or $n = N$), an example is shown in Fig. 2. This means that in the constant force ensemble the tadpole state is unstable, and unfolding of the globule occurs as a globule – stretched chain transition.

**Parameters of the theory**

The model described above depends on the (partial) parameters $\varphi$ (polymer density within the globule), $\mu$ (monomer chemical potential within the globule), and $\gamma$ (interfacial tension coefficient) characterizing the globular state. The values of these parameters are not independent, they are determined by the solvent quality ($\chi$) and chain length (*N*). However, for large globules ($N \gg 1$) $\mu$, $\varphi$, and $\gamma$ are virtually *N*-independent and can be found as functions of the solvent quality only. In Ref. [8] we calculated the values of $\mu$, $\varphi$, and $\gamma$ from SCF modeling of free globules; they are presented in Table 1. The value of the elastic constant $k$ is $k = 0.75$ in accordance with our previous works [7, 8].

|   | $\chi=0.8$ | $\chi=1.0$ | $\chi=1.2$ | $\chi=1.4$ | $\chi=1.6$ | $\chi=1.8$ | $\chi=2.0$ |
|---|---|---|---|---|---|---|---|
| $\mu$ | -0.10 | -0.23 | -0.37 | -0.54 | -0.71 | -0.89 | -1.08 |
| $\varphi$ | 0.54 | 0.70 | 0.80 | 0.87 | 0.92 | 0.95 | 0.96 |
| $\gamma$ | 0.087 | 0.18 | 0.27 | 0.38 | 0.48 | 0.58 | 0.67 |

Table 1. Values of monomer chemical potential ($\mu$) polymer density ($\varphi$) and interfacial tension coefficient ($\gamma$) calculated for different $\chi$ using SCF approach [8].

# Results

## Free energy and deformation curves in *f*-ensemble

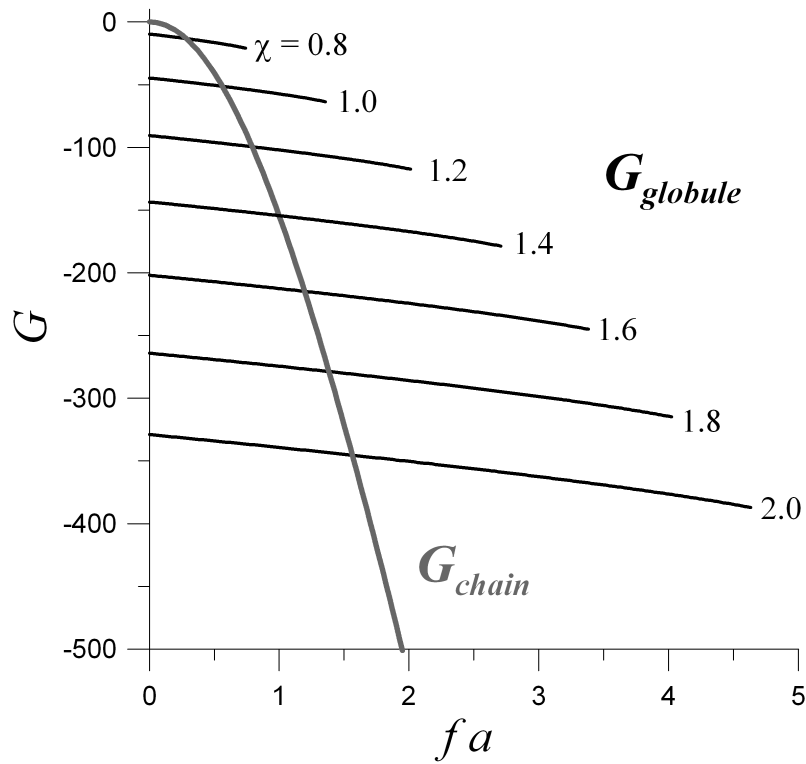

Fig. 3 . Gibbs free energy of compact globular and stretched states as function of applied force for *N*=500 and various values of $\chi$.

Fig. 3 shows dependences of the Gibbs free energies of globular and open chain states on the applied stretching force *f* for *N*=500 and a series of $\chi$ values. Notice that the Gibbs free energy of the open chain $G_{chain}$ is independent on the solvent quality and, therefore is represented by a single curve whereas $G_{globule}$ curves do depend on $\chi$ and terminate at spinodal points $f = f^*$ according to Eq. (8). We see that at small applied forces $G_{globule} < G_{chain}$, with an increase in *f* the difference $G_{chain} - G_{globule}$ decreases and at certain value of *f*, $f = f_{tr}$, two curves intersect and $G_{chain} = G_{globule}$. Then, at $f > f_{tr}$, $G_{globule} > G_{chain}$ and the globule is in open state. In other words, at $f = f_{tr}$ the transition between two states occurs and the globule unfolds. The position of the transition point shifts to the right as the $\chi$ value increases.

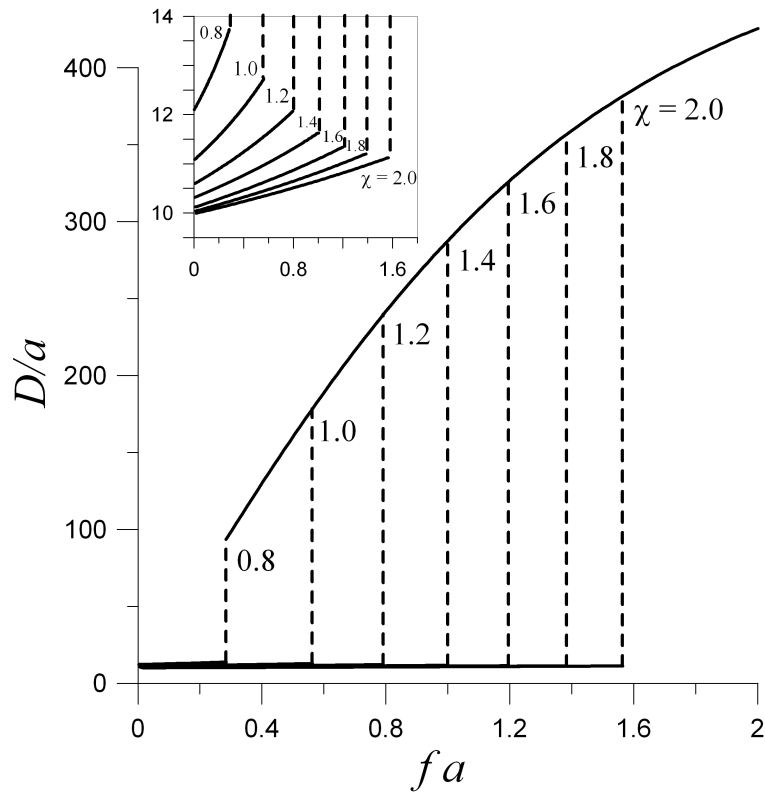

Fig. 4 . Equilibrium extension vs applied force curves for the globule with $N$=500 at various values of $\chi$.

The corresponding $D(f)$ dependences in equilibrium are shown in Fig. 4. At small deformation the end-to end distance increases only weakly, this growth is stronger for better solvent (smaller $\chi$). At the transition point the end-to-end distance abruptly changes, for the given chain length $N = 500$ it is increased by 10 to 50 times. The jump is larger for poorer solvent.

**Comparison of force-extension curves in $D$- and $f$-ensembles**

It is interesting to compare the obtained extension-force dependence with the force-extension relation for the constant extension ensemble obtained in [8]. It can be easily done by exchanging $f$- and $D$- axes in Fig. 4. The result of the comparison is presented in Fig. 5, where force-extension curves for $D$- and $f$- ensembles are shown by solid and dashed lines, respectively. Inspection of the force-extension curves for the $D$- ensemble clearly reveals three different deformation regimes. (1) At small deformations the globule has the shape of an elongated ellipsoid, Fig. 1 b and the reaction force strongly increases with extension.

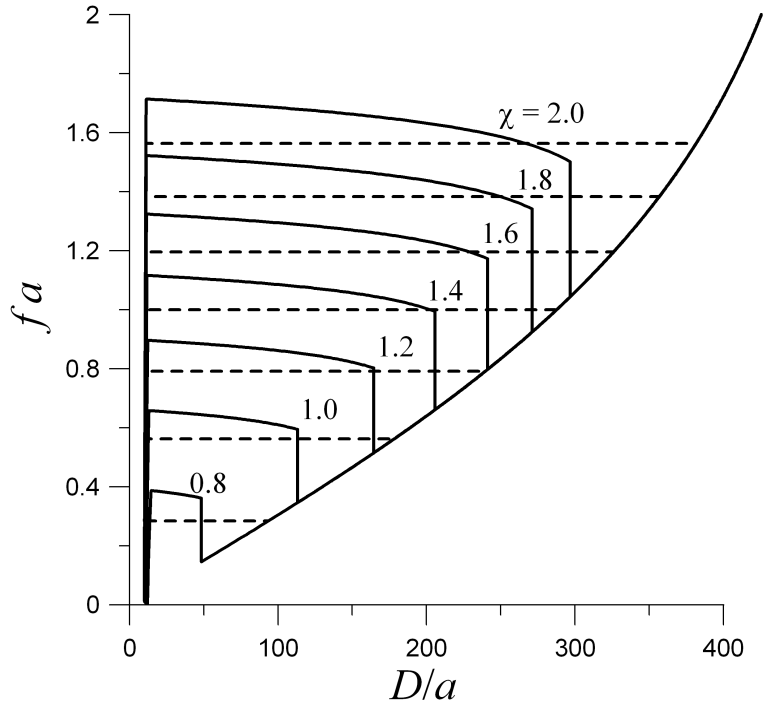

Fig. 5 Force vs extension curves for *N*=500 and various values of calculated in fixed extension (solid curves) and fixed force (dashed curves) ensembles.

(2) At moderate extensions the intramolecular microphase segregation occurs and the globule acquires the tadpole conformation, Fig. 1 c. A progressive increase in the end-to-end distance of the chain is accompanied by a systematic depletion of the globular head and a repartitioning of the monomer units into the stretched tail. At the same time the reaction force slightly decreases (quasi-plateau) as $D$ grows and then drops down – this corresponds to the complete unfolding of the globule.. (3) After the jump, at strong deformations, the globule as a whole is now in the open state, Fig. 1 d, and the reaction force starts to grow again.

Comparing the force-extension curves in the *f*- and *D*-ensembles we see that at weak and strong extensions corresponding to prolate globule and open chain conformations, respectively, the curves coincide, differences are observed at the intermediate extension range. The most important qualitative difference is that the flat "plateau" in the *f*-ensemble does not correspond to the any microphase-segregated state, in contrast to the *D*-ensemble, it only marks the position of the threshold force (transition point). It also follows that not only the microphase segregated tadpole state but also some of the pure states that are stable in the constant extension ensemble cannot be accessed in the *f*-ensemble. This concerns "weakly stretched open chains" (just after the tadpole-open chain [i.e. complete unfolding] transition point in the *D*-ensemble) and "strongly extended globules" (close to the ellipsoid-tadpole transition point in the *D*-ensemble).

**Transition points and phase diagrams in the *f*-ensemble**

Let us consider in more detail the transition point where the abrupt unfolding of the globule occurs. Fig. 6 demonstrates the dependence of the threshold force $f_{tr}$ on the chain length $N$ at different values of $\chi$. One can see that with an increase in both $N$ and/or $\chi$ the value of $f_{tr}$ increases, this growth is more pronounced at smaller $N$.

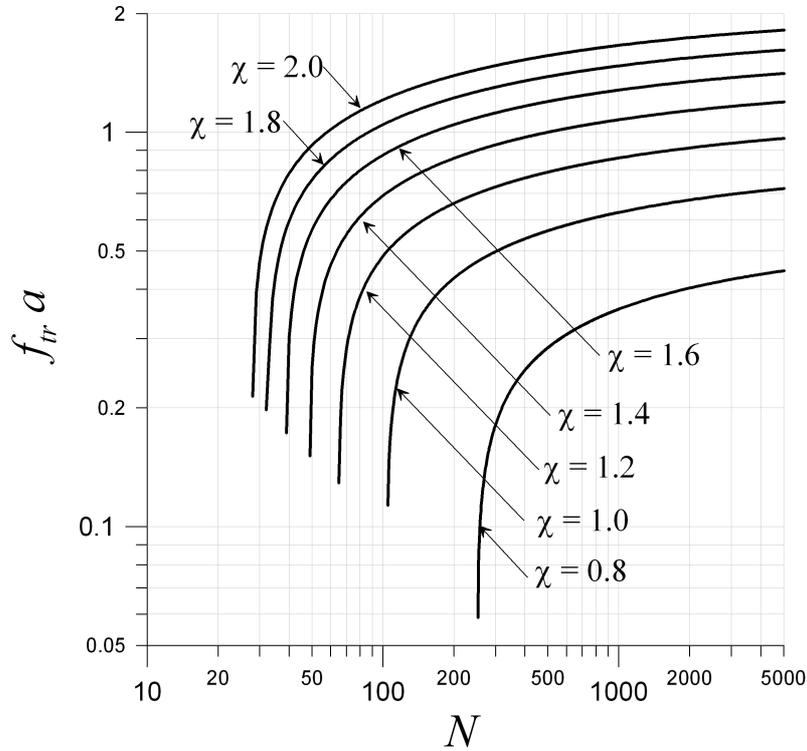

Fig. 6. Critical (transition) force in the fixed force ensemble as function of polymerization degree at various values of $\chi$.

At the transition point the end-to end distance jumps from $D = D_{globule}$ to $D = D_{chain} > D_{globule}$. Fig. 7 shows *N*-dependences of $D_{globule}$ (lower solid curves) and $D_{chain}$ (upper solid curves) for moderately poor ($\chi = 0.8$) and strong ($\chi = 2$) solvent. One can see that the jump in the end-to-end distance at the unfolding transition increases with an increase in both $N$ and/or $\chi$.

For the sake of comparison, critical extension for ellipsoid-tadpole (lower curves)

and tadpole-open chain (upper curves) transition points for the globule unfolded in the *D*-ensemble are also shown in Fig. 7 by dashed lines. The area within dashed curves corresponds to microphase-segregated tadpole states in the *D*-ensemble. Phase diagram for the *D*-ensemble lies completely within the corresponding diagram for the *f*-ensemble. The corresponding small "differences" are related to the above discussed pure states which are stable in the *D*-ensemble but unstable in the *f*-ensemble. These differences, however, decrease as $N$ increases and vanish in the $N \to \infty$ limit.

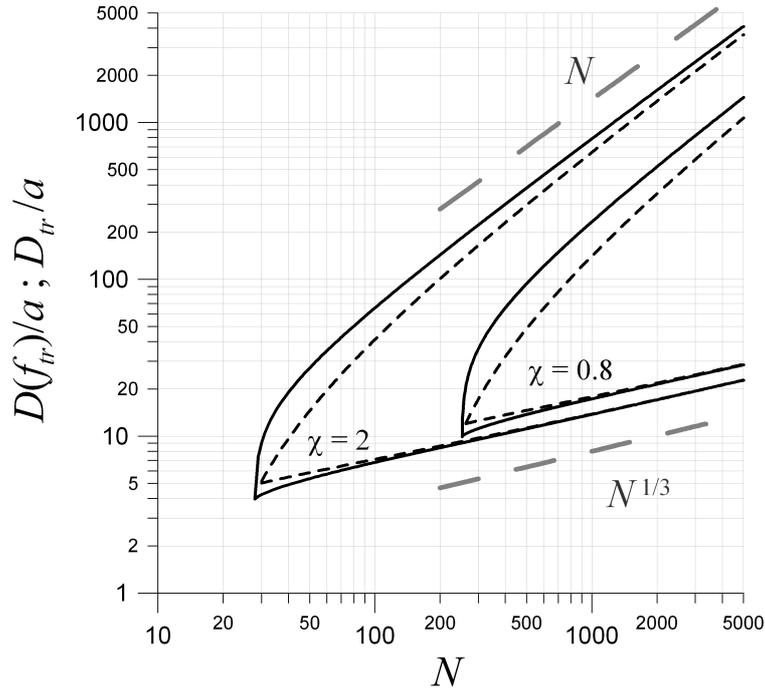

Fig. 7. Globule and open chain extension at critical (transition) force in the *f*-ensemble (solid curves) and critical extensions in the *D*-ensemble (dashed lines) as functions of polymerization degree.

**"Critical point"**

At small $N$, $f_{tr}(N)$ dependences terminate at a certain value of $N = N^*(\chi)$, as it can be seen from Fig. 6. At the same time, in Fig. 7, as $N$ approaches $N^*$ two branches corresponding to $D_{globule}$ and $D_{chain}$ approach each other at $N=N^*$ $D_{globule} = D_{chain}$. Hence, $N^*(\chi)$ can be considered as a "critical point". We are not allowed to call this point as a true critical point (without quotes) because the model employed in the present work is based upon the assumption that the globule is characterized by a constant polymer density and a sharp globule-solvent interface. This simplification can be acceptable for relatively long chain, where the width of the interface Δ is much smaller than the size of the globule ($\Delta \ll R_0$). For short chains, Δ has the same order as $R_0$ and the constant density assumption becomes incorrect and we cannot assert that the "critical point" observed in Figs. 6 and 7 is a true critical point.

On the other hand it was demonstrated by SCF modeling of globule unfolding in the constant extension ensemble [7], that even below the critical point the force-extension curve has a characteristic loop (see the force-extension curve for $N=200$, $\chi =0.8$ in Figure 4 of Ref. [7]) which is a sign of the instability and jumpwise unfolding transition in the conjugate *f*-ensemble. We can anticipate that a true critical point should be shifted towards smaller *N*. This length should be related to the minimal chain length necessary for the globule formation.

**Conclusions**

In conclusion, we studied theoretically the unfolding of a homopolymer globule by applied force (in the *f*-ensemble). We have shown that an increase in the pulling force leads to abrupt unfolding of the globule. The transition is of "all or none" type, i.e. no mixed (microphase segregated) state appears, and is accompanied by a jump in the end-to-end distance. The values of the threshold force and of the jump grow with an increase in *N* and/or χ. We also compared globule deformation in *f*- and *D*-ensembles and shown that they are not equivalent for a finite *N*.

Finally, it is important to note that the developed theory is essentially equilibrium one. It assumes that at each applied force *f* or imposed deformation *D* the system has enough time to go over energetic barrier(s) and reach an equilibrium state. However, locally stable intermediate states, where the system can be trapped, may play an essential role in determining unfolding pathway(s) and characteristic time (see, for example, recent experimental work [9]).

Acknnowledgements: This work is supported by Russian Foundation for Basic Research (grant 08-03-00336-a).